\documentclass[conference,a4paper]{IEEEtran}

\normalsize

\ifCLASSINFOpdf
\else
\fi

\usepackage{amsmath}
\usepackage{amssymb}
\usepackage{cite}
\usepackage{graphicx}
\usepackage{algorithm}
\usepackage{algpseudocode}
\usepackage{subfigure}
\usepackage{multirow} 
\usepackage{longtable}
\usepackage{threeparttable}
\usepackage{caption3}
\usepackage{array}
\usepackage{cases}
\usepackage{bbm}
\usepackage{color}
\usepackage[T1]{fontenc}
\usepackage[utf8]{inputenc}

\newcommand\mycom[2]{\genfrac{}{}{0pt}{}{#1}{#2}}

\begin{document}
\title{Waveform Optimization for SWIPT with Nonlinear Energy Harvester Modeling}

\author{\IEEEauthorblockN{Bruno Clerckx}
\IEEEauthorblockA{Communication and Signal Processing Group, EEE Department, Imperial College London, United Kingdom\\
School of Electrical Engineering, Korea University, Korea\\
Email: b.clerckx@imperial.ac.uk}}

\maketitle

\begin{abstract} Simultaneous Wireless Information and Power Transfer (SWIPT) has attracted significant attention in the communication community. The problem of waveform design for SWIPT has however never been addressed so far. In this paper, a novel SWIPT transceiver architecture is introduced relying on the superposition of multisine and OFDM waveforms at the transmitter and a power-splitter receiver equipped with an energy harvester and an information decoder capable of cancelling the multisine waveforms. The SWIPT multisine/OFDM waveforms are optimized so as to maximize the rate-energy region of the whole system. They are adaptive to the channel state information and result from a posynomial maximization problem that originates from the non-linearity of the energy harvester. Numerical results illustrate the performance of the derived waveforms and SWIPT architecture.
\footnote{This work has been partially supported by the EPSRC of the UK under grant EP/M008193/1.}
\end{abstract}


\IEEEpeerreviewmaketitle

\section{Introduction}


\par Simultaneous Wireless Information and Power Transfer (SWIPT) has recently attracted significant attention in academia, with works addressing many scenarios, a.o.\ MIMO broadcasting \cite{Zhang:2013}, architecture design \cite{Zhou:2013}, interference channel \cite{Park:2013,Park:2014}, broadband systems \cite{Huang:2013}, relaying \cite{Nasir:2013,Huang:2015}. 
\par The core element of the SWIPT receiver that enables to harvest wireless energy is the rectenna. The rectenna is made of a non-linear device followed by a low-pass filter to extract a DC power out of an RF input signal. The amount of DC power collected is a function of the input power level and the RF-to-DC conversion efficiency. Interestingly, the RF-to-DC conversion efficiency is not only a function of the rectenna design but also of its input waveform \cite{Trotter:2009,Boaventura:2011,Collado:2014,Clerckx:2015}.
\par In the rapidly expanding SWIPT literature, the sensitivity of the RF-to-DC conversion efficiency to the rectenna design and input waveforms has been inaccurately addressed in past SWIPT works (e.g. \cite{Zhang:2013,Zhou:2013,Park:2013,Park:2014,Huang:2013,Nasir:2013,Huang:2015}). It is indeed assumed for the sake of simplicity and tractability that the harvested DC power is modeled as a conversion efficiency constant multiplied by the average power of the input signal to the energy harvester. Unfortunately, this is an oversimplified model that does not reflect accurately the dependence w.r.t.\ the input waveform. This inaccuracy originates from the truncation to the second order of the non-linear rectification process of the diode \cite{Ladan:2015}. Hence, truncating to a second order the non-linear rectification process of the diode has been used so far so as to simplify the design of SWIPT but is unfortunately an unrealistic assumption from an RF perspective \cite{Trotter:2009,Boaventura:2011,Collado:2014,Ladan:2015} that can lead to inaccurate or inefficient design of SWIPT.
\par The design of SWIPT waveform that accounts for the rectifier non-linearity has never been addressed so far. However, since SWIPT relies on WPT, a thorough understanding of the WPT waveform design would be required beforehand. In \cite{Clerckx:2015}, the WPT waveform design problem has been tackled by introducing a tractable analytical model of the non-linearity of the diode through the second and fourth order terms in the Taylor expansion of the diode characteristics. Assuming perfect Channel State Information at the Transmitter (CSIT), an optimization problem was formulated to adaptively change on each transmit antenna a multisine waveform as a function of the CSI so as to maximize the rectifier output DC current. Significant performance gains of the optimized waveforms over state-of-the-art waveforms were demonstrated.
\par In this paper we leverage the waveform optimization for WPT in \cite{Clerckx:2015} and tackle the problem of waveform and transceiver optimization for Multiple Input-Single Output (MISO) SWIPT. A novel SWIPT transceiver architecture is introduced relying on the superposition of multisine waveforms for WPT and OFDM waveforms for Wireless Information Transfer (WIT) at the transmitter and a power-splitter receiver equipped with an energy harvester and an information decoder capable of cancelling the multisine waveforms. The SWIPT multisine/OFDM waveforms are optimized so as to maximize the rate-energy region of the whole system, accounting for the non-linearity of the energy harvester.   

\par \textit{Organization:} Section \ref{SWIPT_section} introduces the SWIPT architecture, section \ref{section_SWIPT_waveform} addresses the SWIPT waveform design, section \ref{simulations} evaluates the performance and section \ref{conclusions} concludes the work.
\par \textit{Notations:} Bold lower case and upper case letters stand for vectors and matrices respectively whereas a symbol not in bold font represents a scalar. $\left\|.\right\|_F^2$ refers to the Frobenius norm a matrix. $\mathcal{A}\left\{.\right\}$ refers to the DC component of a signal. $\mathcal{E}_X\left\{.\right\}$ refers to the expectation operator taken over the distribution of the random variable $X$ ($X$ may be omitted for readability if the context is clear). 
$\left(.\right)^T$ and $\left(.\right)^H$ represent the transpose and conjugate transpose of a matrix or vector respectively. 

\section{A SWIPT Transceiver Architecture}\label{SWIPT_section}
\begin{figure}
\begin{minipage}[c]{\columnwidth}
\centering
\includegraphics[width=0.45\columnwidth]{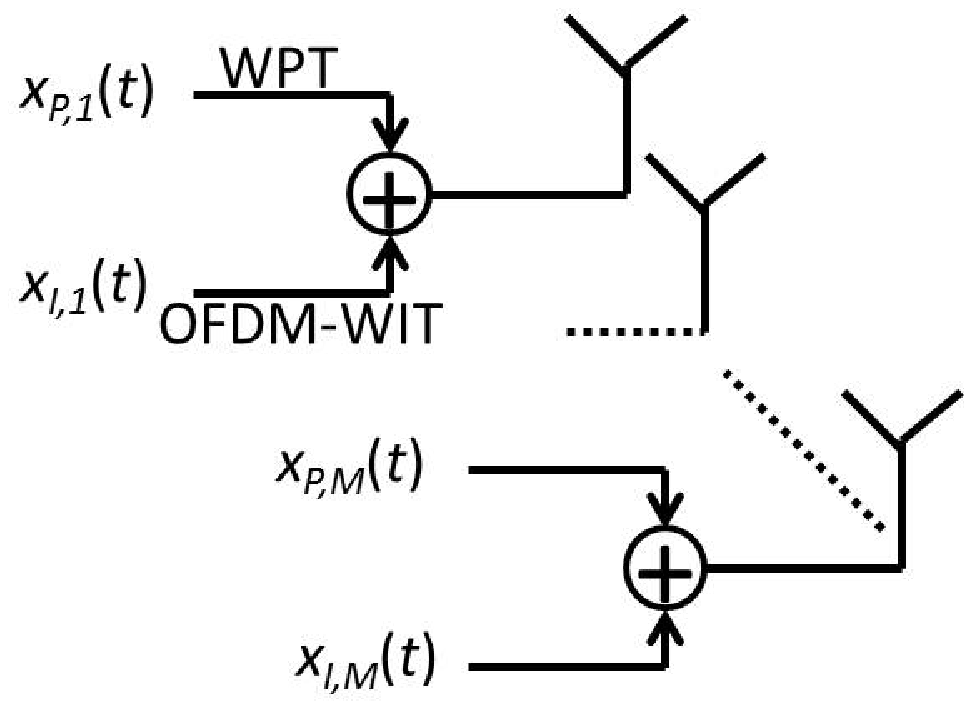}\\
\small{(a) Transmitter}
\end{minipage}\vspace{0.2cm}
\begin{minipage}[c]{\columnwidth}
\centering
\includegraphics[width=0.85\columnwidth]{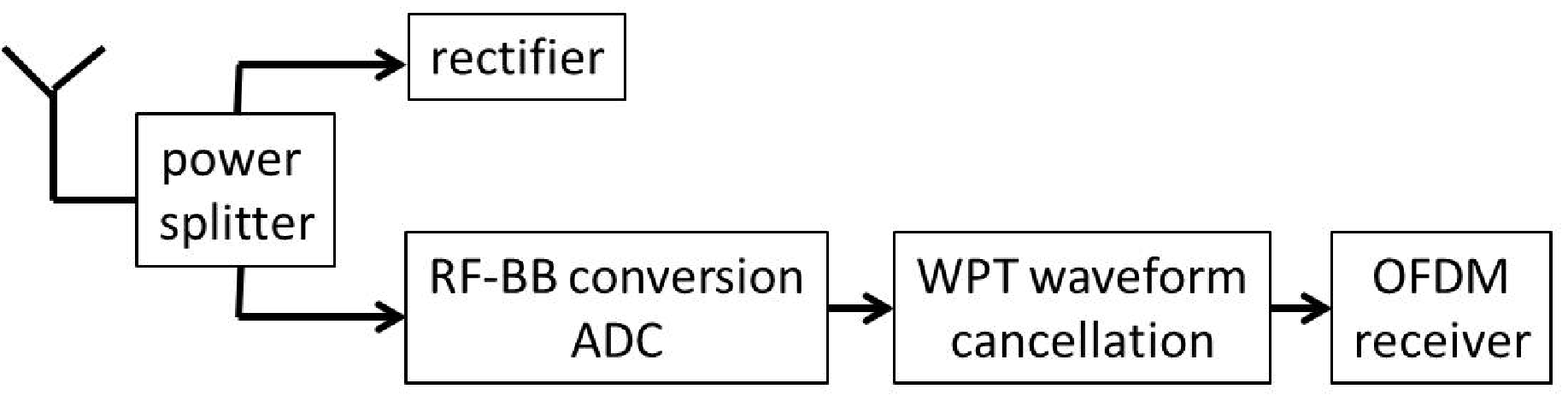}\\
\small{(b) Receiver}
\end{minipage}
\caption{A transceiver architecture for SWIPT.}
\label{WIPT_transceiver}
\end{figure}

In Figure \ref{WIPT_transceiver}, we introduce a SWIPT architecture where power and information are transmitted simultaneously from one transmitter to one receiver equipped with a power splitter.

\subsection{Transmitter}
The SWIPT waveform on antenna $m$, $x_{m}(t)$, consists in the superposition of one multisine waveform $x_{P,m}(t)$ at frequencies $w_n=w_0+n\Delta_w$ (with $\Delta_w=2\pi\Delta_f$ the frequency spacing), $n=0,\ldots,N-1$ for WPT and one OFDM waveform $x_{I,m}(t)$ at the same frequencies for WIT. 
\par The multisine WPT waveform is written as 
\begin{equation}
x_{P,m}(t)=\sum_{n=0}^{N-1} s_{P,n,m} \cos(w_n t+\phi_{P,n,m}).
\end{equation}
\par The baseband OFDM signal over one symbol duration $T=1/\Delta_f$ is written as $x_{B,m}(t)=\sum_{n=0}^{N-1} x_{n,m} e^{j\frac{2\pi t}{T}n}$, $0\leq t \leq T$,
where $x_{n,m}=w_{I,n,m}\tilde{x}_{n}$ refers to the precoded input symbol on tone $n$ and antenna $m$. We further write the precoder $w_{I,n,m}=\left|w_{I,n,m}\right|e^{j\phi_{I,n,m}}$ and the input symbol $\tilde{x}_{n}=\left|\tilde{x}_{n}\right|e^{j\phi_{\tilde{x}_n}}$.
After adding the cyclic prefix over duration $T_g$, $x_{B,m}(t)=\sum_{n=0}^{N-1} x_{n,m} e^{j\frac{2\pi t}{T}n}$, $-T_g\leq t \leq T$.
Vector-wise, the baseband OFDM signal vector writes as $\mathbf{x}_{B}(t)=\left[\begin{array}{ccc}x_{B,1}(t) & \ldots & x_{B,M}(t)\end{array}\right]^T=\sum_{n=0}^{N-1} \mathbf{x}_{n} e^{j\frac{2\pi t}{T}n}$
with $\mathbf{x}_{n}=\mathbf{w}_{I,n}\tilde{x}_{n}$ and $\mathbf{w}_{I,n}=\left[\begin{array}{ccc}w_{I,n,1}&\ldots & w_{I,n,M}\end{array}\right]^T$ is the precoder.
After upconversion, the transmit OFDM signal on antenna $m$ is written as
\begin{align}
x_{I,m}(t)&=\Re\left\{x_{B,m}(t)e^{j w_0 t}\right\}\nonumber\\
&=\sum_{n=0}^{N-1} \tilde{s}_{I,n,m} \cos(w_n t+\tilde{\phi}_{I,n,m})
\end{align}
where $x_{n,m}=\tilde{s}_{I,n,m}e^{j\tilde{\phi}_{I,n,m}}$ with $\tilde{s}_{I,n,m}=\left|w_{I,n,m}\right|\left|\tilde{x}_{n}\right|$ and $\tilde{\phi}_{I,n,m}=\phi_{I,n,m}+\phi_{\tilde{x}_n}$. We also define $s_{I,n,m}=\sqrt{P_{I,n}}\left|w_{I,n,m}\right|$ where $P_{I,n}=\mathcal{E}\big\{\left|\tilde{x}_n\right|^2\big\}$.
\par From Fig \ref{WIPT_transceiver}(a), the SWIPT waveform on antenna $m$ is
\begin{align}
x_{m}(t)&=\sum_{n=0}^{N-1} s_{P,n,m} \cos(w_n t+\phi_{P,n,m})\nonumber\\
&\hspace{1cm}+\tilde{s}_{I,n,m} \cos(w_n t+\tilde{\phi}_{I,n,m}).
\end{align}

\par The amplitudes and phases of the WPT waveform are collected into $N \times M$ matrices $\mathbf{S}_P$ and $\mathbf{\Phi}_P$, respectively. Similarly, the $(n,m)$ entry of matrix $\tilde{\mathbf{S}}_I$, $\mathbf{S}_I$, $\mathbf{\Phi}_I$ write as $\tilde{s}_{I,n,m}$, $s_{I,n,m}$, $\phi_{I,n,m}$, respectively.
We define the average power of the WPT and WIT waveforms as $P_P=\frac{1}{2}\left\|\mathbf{S}_P\right\|_F^2$ and $P_I=\frac{1}{2}\mathcal{E}\big\{\big\|\tilde{\mathbf{S}}_I\big\|_F^2\big\}=\frac{1}{2}\left\|\mathbf{S}_I\right\|_F^2$. The total average transmit power constraint writes as $P_P+P_I\leq P$. 

\subsection{Receiver}

\par The multi-antenna transmitted sinewaves propagate through a multipath channel, characterized by $L$ paths whose delay, amplitude, phase and direction of departure (chosen with respect to the array axis) are respectively denoted as $\tau_l$, $\alpha_l$, $\xi_l$ and $\theta_l$, $l=1,\ldots,L$. We assume transmit antennas are closely located so that $\tau_l$, $\alpha_l$ and $\xi_l$ are the same for all transmit antennas (assumption of a narrowband balanced array) \cite{Clerckx:2013}. Taking the power signal for instance, it is transmitted by antenna $m$ and received at the single-antenna receiver after multipath propagation as
\begin{align}\label{received_signal_ant_m}
&y_P^{(m)}(t)\\
&=\sum_{n=0}^{N-1} s_{P,n,m}\left(\sum_{l=0}^{L-1}\alpha_l \cos(w_n(t-\tau_l)+\xi_l+\phi_{P,n,m}+\Delta_{n,m,l})\right)\nonumber
\end{align}
where $\Delta_{n,m,l}$ refers to the phase shift between the $m^{th}$ transmit antenna and the first one. For simplicity, we assume that $\Delta_{n,1,l}=0$. For a Uniform Linear Array (ULA), $\Delta_{n,m,l}=2\pi (m-1)\frac{d}{\lambda_n}\cos(\theta_l)$ where $d$ is the inter-element spacing, $\lambda_n$ the wavelength of the $n^{th}$ sinewave.

\par The quantity between the brackets in \eqref{received_signal_ant_m} can simply be rewritten as 
\begin{multline}
\sum_{l=0}^{L-1}\alpha_l \cos(w_n(t-\tau_l)+\xi_l+\phi_{P,n,m}+\Delta_{n,m,l})\\
=A_{n,m} \cos(w_n t+\psi_{P,n,m})
\end{multline}
where the amplitude $A_{n,m}$ and the phase $\psi_{P,n,m}$ are such that
\begin{align}\label{A_nm_psi}
A_{n,m}e^{j \psi_{P,n,m}}&=A_{n,m}e^{j \left(\phi_{P,n,m}+\bar{\psi}_{n,m}\right)}=e^{j \phi_{P,n,m}}h_{n,m}
\end{align}
with $h_{n,m}=A_{n,m}e^{j \bar{\psi}_{n,m}}=\sum_{l=0}^{L-1}\alpha_l e^{j(-w_n\tau_l+\Delta_{n,m,l}+\xi_l)}$ the frequency response of the channel of antenna $m$ at $w_n$. Vector-wise, we can define the frequency-domain channel vector $\mathbf{h}_n=\big[\begin{array}{ccc}h_{n,1}&\ldots & h_{n,M}\end{array}\big]$. We can write similar expressions for the information signal.

\par At the receiver, we can write the received signal as $y(t)=y_{P}(t)+y_I(t)$, i.e.\ the sum of two contributions at the output of the channel, namely one from WPT $y_{P}(t)$ and the other from WIT $y_{I}(t)$
\begin{align}
y_P(t)&=\sum_{m=1}^{M}\sum_{n=0}^{N-1}s_{P,n,m}A_{n,m} \cos(w_n t+\psi_{P,n,m})\\
y_I(t)&=\sum_{m=1}^{M}\sum_{n=0}^{N-1}\tilde{s}_{I,n,m}A_{n,m} \cos(w_n t+\tilde{\psi}_{I,n,m})
\end{align}
where $\psi_{P,n,m}=\phi_{P,n,m}+\bar{\psi}_{n,m}$ and $\tilde{\psi}_{I,n,m}=\tilde{\phi}_{I,n,m}+\bar{\psi}_{n,m}=\phi_{I,n,m}+\phi_{\tilde{x}_n}+\bar{\psi}_{n,m}$. Let us also define $\psi_{I,n,m}=\phi_{I,n,m}+\bar{\psi}_{n,m}$ such that $\tilde{\psi}_{I,n,m_0}-\tilde{\psi}_{I,n,m_1}=\psi_{I,n,m_0}-\psi_{I,n,m_1}$. Using a power splitter with a power splitting ratio $\rho$ and assuming perfect matching (as in \cite{Clerckx:2015}), the input voltage signal $\sqrt{\rho R_{ant}}y(t)$ is conveyed to the input to the energy harvester (EH) while $\sqrt{(1-\rho)R_{ant}}y(t)$ is conveyed to the information decoder (ID). 

\subsubsection{ID receiver} Since $x_{P,m}(t)$ does not contain any information, it is deterministic and can be cancelled at the ID receiver. Therefore, after down-conversion and ADC, the contribution of the WPT waveform is subtracted from the received signal (Figure \ref{WIPT_transceiver}(b)). Conventional OFDM processing is then conducted, namely removing the cyclic prefix and performing FFT. We can write the equivalent baseband system model of the ID receiver as
\begin{equation}
y_{ID,n}=\sqrt{1-\rho}\mathbf{h}_n\mathbf{w}_{I,n}\tilde{x}_n+v_n
\end{equation}
where $v_n$ is the AWGN noise on tone $n$ (with variance $\sigma_n^2$) originating from the antenna and the RF to baseband down-conversion. 

\par Assuming perfect cancellation and complex Gaussian input symbols $\left\{\tilde{x}_n\right\}$, the rate writes as
\begin{equation}
I(\mathbf{S}_I,\mathbf{\Phi}_I,\rho)=\sum_{n=0}^{N-1} \log_2\left(1+\frac{(1-\rho)P_{I,n}}{\sigma_n^2} \left|\mathbf{h}_n\mathbf{w}_{I,n}\right|^2\right).\label{R}
\end{equation} 
Naturally, $I(\mathbf{S}_I,\mathbf{\Phi}_I,\rho)$ can never be larger than the maximum rate achievable when $\rho=0$, i.e. $I(\mathbf{S}_I^{\star},\mathbf{\Phi}_I^{\star},0)$, which is obtained by performing matched filtering on each subcarrier and water-filling power allocation across subcarrier.

\subsubsection{EH receiver} At the energy harvester, following \cite{Clerckx:2015}, the DC component of the current at the output of the rectifier is proportional to the quantity $z_{DC}=k_2 \rho R_{ant}\mathcal{A}\left\{y(t)^2\right\}+k_4 R_{ant}^2 \rho^2\mathcal{A}\left\{y(t)^4\right\}$ where $R_{ant}$ is the antenna impedance and $k_i=i_s\frac{e^{\frac{a}{n v_t}}}{i!\left(n v_t\right)^i}$, $i=2,4$. Contrary to WPT, in SWIPT, both WPT and WIT now contribute to the DC component $z_{DC}$. For a given channel impulse response, the input symbols $\left\{\tilde{x}_n\right\}$ change randomly every symbol duration $T$. The DC component $z_{DC}$ therefore needs to be averaged out over the distribution of the input symbols $\left\{\tilde{x}_n\right\}$ such that $z_{DC}=\mathcal{E}_{\left\{\tilde{x}_n\right\}}\left\{k_2 \rho R_{ant}\mathcal{A}\left\{y(t)^2\right\}+k_4 R_{ant}^2 \rho^2\mathcal{A}\left\{y(t)^4\right\}\right\}$. This enables to compute the DC component as in \eqref{z_DC_SWIPT}, where we use the fact that $\mathcal{E}\left\{\mathcal{A}\left\{y_P(t)y_I(t)\right\}\right\}=0$, $\mathcal{E}\left\{\mathcal{A}\left\{y_P(t)^3y_I(t)\right\}\right\}=0$, $\mathcal{E}\left\{\mathcal{A}\left\{y_P(t)y_I(t)^3\right\}\right\}=0$ and $\mathcal{E}\left\{\mathcal{A}\left\{y_P(t)^2y_I(t)^2\right\}\right\}=\mathcal{A}\left\{y_P(t)^2\right\}\mathcal{E}\left\{\mathcal{A}\left\{y_I(t)^2\right\}\right\}$.
\begin{table*}
\begin{multline}\label{z_DC_SWIPT}
z_{DC}(\mathbf{S}_P,\mathbf{S}_I,\mathbf{\Phi}_P,\mathbf{\Phi}_I,\rho)=k_2 \rho R_{ant}\mathcal{A}\left\{y_P(t)^2\right\}+k_4\rho^2 R_{ant}^2\mathcal{A}\left\{y_P(t)^4\right\}+k_2 \rho R_{ant}\mathcal{E}\left\{\mathcal{A}\left\{y_I(t)^2\right\}\right\}\\
+k_4\rho^2 R_{ant}^2 \mathcal{E}\left\{\mathcal{A}\left\{y_I(t)^4\right\}\right\}+6k_4\rho^2 R_{ant}^2 \mathcal{A}\left\{y_P(t)^2\right\}\mathcal{E}\left\{\mathcal{A}\left\{y_I(t)^2\right\}\right\}.
\end{multline} 
\hrulefill
\end{table*}

Quantities $\mathcal{A}\left\{y_P(t)^2\right\}$ and $\mathcal{A}\left\{y_P(t)^4\right\}$ can be directly obtained from the WPT expressions in \cite{Clerckx:2015} and reproduced in \eqref{E_y_P_2} and \eqref{E_y_P_4} for simplicity.
For $\mathcal{E}\left\{\mathcal{A}\left\{y_I(t)^2\right\}\right\}$ and $\mathcal{E}\left\{\mathcal{A}\left\{y_I(t)^4\right\}\right\}$, the DC component is first extracted for a given set of amplitudes $\left\{\tilde{s}_{I,n,m}\right\}$ and phases $\big\{\tilde{\phi}_{I,n,m}\big\}$ and then expectation is taken over the randomness of the input symbols $\tilde{x}_n$. Due to the complex Gaussian distribution of the input symbols, $\left|\tilde{x}_n\right|^2$ is exponentially distributed with $\mathcal{E}\left\{\big|\tilde{x}_n\right|^2\big\}=P_{I,n}$ and $\phi_{\tilde{x}_n}$ is uniformly distributed. From the moments of an exponential distribution, we also have that $\mathcal{E}\left\{\big|\tilde{x}_n\right|^4\big\}=2P_{I,n}^2$. This helps expressing \eqref{E_y_I_2} and \eqref{E_y_I_4} as a function of $s_{I,n,m}=\sqrt{P_{I,n}}\left|w_{I,n,m}\right|$. Note that the rectenna harvests energy from the superposed waveform but the contribution of each waveform to $z_{DC}$ is different given the different nature of the waveforms (WPT is deterministic while WIT exhibits some randomness due to information) and the non-linearity of the rectenna.
\begin{table*}
\begin{align} 
\mathcal{A}\left\{y_P(t)^2\right\}&=\frac{1}{2}\left[\sum_{n=0}^{N-1} \sum_{m_0,m_1} s_{P,n,m_0}s_{P,n,m_1}A_{n,m_0}A_{n,m_1}\cos\left(\psi_{P,n,m_0}-\psi_{P,n,m_1}\right)\right]\label{E_y_P_2}\\
\mathcal{A}\left\{y_P(t)^4\right\}&=\frac{3}{8}\left[\sum_{\mycom{n_0,n_1,n_2,n_3}{n_0+n_1=n_2+n_3}}\sum_{\mycom{m_0,m_1,}{m_2,m_3}}\Bigg[\prod_{j=0}^3s_{P,n_j,m_j}A_{n_j,m_j}\Bigg]\cos(\psi_{P,n_0,m_0}+\psi_{P,n_1,m_1}-\psi_{P,n_2,m_2}-\psi_{P,n_3,m_3})\right]\label{E_y_P_4}\\
\mathcal{E}\left\{\mathcal{A}\left\{y_I(t)^2\right\}\right\}
&=\frac{1}{2}\left[\sum_{n=0}^{N-1} \sum_{m_0,m_1} s_{I,n,m_0}s_{I,n,m_1}A_{n,m_0}A_{n,m_1}\cos\left(\psi_{I,n,m_0}-\psi_{I,n,m_1}\right)\right]
\label{E_y_I_2}\\
\mathcal{E}\left\{\mathcal{A}\left\{y_I(t)^4\right\}\right\}
&=\frac{6}{8}\Bigg[\sum_{n_0,n_1}\sum_{\mycom{m_0,m_1,}{m_2,m_3}}\Bigg[\prod_{j=0,2}s_{I,n_0,m_j}A_{n_0,m_j}\Bigg]\Bigg[\prod_{j=1,3}s_{I,n_1,m_j}A_{n_1,m_j}\Bigg]
\cos(\psi_{I,n_0,m_0}+\psi_{I,n_1,m_1}-\psi_{I,n_0,m_2}-\psi_{I,n_1,m_3})
\Bigg]
\label{E_y_I_4}
\end{align}
\hrulefill
\end{table*}

\section{SWIPT Waveform Optimization}\label{section_SWIPT_waveform}
We can now define the achievable rate-harvested energy (or more accurately rate-DC current) region as
\begin{multline}
C_{R-I_{DC}}(P)\triangleq\Big\{(R,I_{DC}):R\leq I(\mathbf{S}_I,\mathbf{\Phi}_I,\rho),\Big. \\
\Big.I_{DC}\leq z_{DC}(\mathbf{S}_P,\mathbf{S}_I,\mathbf{\Phi}_P,\mathbf{\Phi}_I,\rho), \frac{1}{2}\big[\left\|\mathbf{S}_I\right\|_F^2+\left\|\mathbf{S}_P\right\|_F^2\big]\leq P \Big\}.
\end{multline}
Optimal values $\mathbf{S}_P^{\star}$,$\mathbf{S}_I^{\star}$,$\mathbf{\Phi}_P^{\star}$,$\mathbf{\Phi}_I^{\star},\rho^{\star}$ are to be found in order to enlarge as much as possible the rate-harvested energy region.

\subsection{Phase Optimization}
In order to maximize the rate \eqref{R}, $\mathbf{w}_n$ should be chosen as a transmit matched filter, i.e.\ $\mathbf{w}_n=\mathbf{h}_n^H/\left\|\mathbf{h}_n\right\|$. However, $\mathbf{w}_n$ also influences the amount of DC current $z_{DC}$ and a transmit matched filter may not be a suitable strategy to also maximize $z_{DC}$. Looking at \eqref{R} and \eqref{z_DC_SWIPT}, we can nevertheless conclude that matched filtering w.r.t.\ the phases of the channel is optimal from both rate and harvested energy maximization perspective. This leads to the same phase decisions as for WPT in \cite{Clerckx:2015}, namely $\phi_{P,n,m}^{\star}=\phi_{I,n,m}^{\star}=-\bar{\psi}_{n,m}$ and guarantees all arguments of the cosine functions in $\left\{\mathcal{A}\left\{y_P(t)^i\right\}\right\}_{i=2,4}$ (expressions \eqref{E_y_P_2} and \eqref{E_y_P_4}) and in $\left\{\mathcal{E}\left\{\mathcal{A}\left\{y_I(t)^i\right\}\right\}\right\}_{i=2,4}$ (expressions \eqref{E_y_I_2} and \eqref{E_y_I_4}) to be equal to 0. $\mathbf{\Phi}_P^{\star}$ and $\mathbf{\Phi}_I^{\star}$ are obtained by collecting $\phi_{P,n,m}^{\star}$ and $\phi_{I,n,m}^{\star}$ $\forall n,m$ into a matrix, respectively.

\subsection{Amplitude and Power Split Optimization}

\begin{table*}
\begin{align}\label{z_DC_SWIPT_final}
&z_{DC}(\mathbf{S}_P,\mathbf{S}_I,\mathbf{\Phi}_P^{\star},\mathbf{\Phi}_I^{\star},\rho)\nonumber\\
&\hspace{0.3cm}= \frac{k_2 \rho}{2}R_{ant}\left[\sum_{n=0}^{N-1} \sum_{m_0,m_1} \Bigg[\prod_{j=0}^1 s_{P,n,m_j}A_{n,m_j}\Bigg]\right]
+ \frac{3k_4\rho^2}{8}R_{ant}^2\left[\sum_{\mycom{n_0,n_1,n_2,n_3}{n_0+n_1=n_2+n_3}}\sum_{\mycom{m_0,m_1,}{m_2,m_3}}\Bigg[\prod_{j=0}^3s_{P,n_j,m_j}A_{n_j,m_j}\Bigg]\right]\nonumber\\
&\hspace{0.6cm}+\frac{k_2 \rho}{2}R_{ant}\left[\sum_{n=0}^{N-1} \sum_{m_0,m_1} \Bigg[\prod_{j=0}^1 s_{I,n,m_j}A_{n,m_j}\Bigg]\right]+ \frac{3 k_4\rho^2}{4}R_{ant}^2\left[\sum_{n=0}^{N-1} \sum_{m_0,m_1} \Bigg[\prod_{j=0}^1 s_{I,n,m_j}A_{n,m_j}\Bigg]\right]^2 \nonumber\\
&\hspace{0.6cm}+\frac{3k_4\rho^2}{2} R_{ant}^2\left[\sum_{n=0}^{N-1} \sum_{m_0,m_1} \Bigg[\prod_{j=0}^1 s_{P,n,m_j}A_{n,m_j}\Bigg]\right]\left[\sum_{n=0}^{N-1} \sum_{m_0,m_1} \Bigg[\prod_{j=0}^1 s_{I,n,m_j}A_{n,m_j}\Bigg]\right]
\end{align} 
\hrulefill
\end{table*}
\par With such phases $\mathbf{\Phi}_P^{\star}$ and $\mathbf{\Phi}_I^{\star}$, $z_{DC}(\mathbf{S}_P,\mathbf{S}_I,\mathbf{\Phi}_P^{\star},\mathbf{\Phi}_I^{\star},\rho)$ can be finally written as \eqref{z_DC_SWIPT_final}. Similarly we can write
\begin{equation}
I(\mathbf{S}_I,\mathbf{\Phi}_I^{\star},\rho)=\log_2\left(\prod_{n=0}^{N-1}\left(1+\frac{(1-\rho)}{\sigma_n^2} C_n\right)\right)\label{R_SWIPT_final} 
\end{equation}
where $C_n=\sum_{m_0,m_1} \prod_{j=0}^1 s_{I,n,m_j}A_{n,m_j}$. 

\par Recall from \cite{Chiang:2005} that a monomial is defined as the function $g:\mathbb{R}_{++}^{N}\rightarrow\mathbb{R}:g(\mathbf{x})=c x_1^{a_1}x_2^{a_2}\ldots x_N^{a_N}$
where $c>0$ and $a_i\in\mathbb{R}$. A sum of $K$ monomials is called a posynomial and can be written as $f(\mathbf{x})=\sum_{k=1}^K g_k(\mathbf{x})$ with $g_k(\mathbf{x})=c_k x_1^{a_{1k}}x_2^{a_{2k}}\ldots x_N^{a_{Nk}}$ where $c_k>0$. As we can see from \eqref{z_DC_SWIPT_final}, $z_{DC}(\mathbf{S}_P,\mathbf{S}_I,\mathbf{\Phi}_P^{\star},\mathbf{\Phi}_I^{\star},\rho)$ is a posynomial. 

\par In order to identify the achievable rate-energy region, we formulate the optimization problem as an energy maximization problem subject to transmit power and rate constraints
\begin{align}\label{SWIPT_opt_problem}
\max_{\mathbf{S}_P,\mathbf{S}_I,\rho} \hspace{0.3cm}&z_{DC}(\mathbf{S}_P,\mathbf{S}_I,\mathbf{\Phi}_P^{\star},\mathbf{\Phi}_I^{\star},\rho)\\
\textnormal{subject to} \hspace{0.3cm} &\frac{1}{2}\big[\left\|\mathbf{S}_I\right\|_F^2+\left\|\mathbf{S}_P\right\|_F^2\big]\leq P,\\
& I(\mathbf{S}_I,\mathbf{\Phi}_I^{\star},\rho)\geq \bar{R}.
\end{align}
It therefore consists in maximizing a posynomial subject to constraints. 
Unfortunately this problem is not a standard Geometric Program (GP) but it can be transformed to an equivalent problem by introducing an auxiliary variable $t_0$
\begin{align}\label{SWIPT_opt_problem_eq}
\min_{\mathbf{S}_P,\mathbf{S}_I,\rho,t_0} \hspace{0.3cm} &1/t_0\\
\textnormal{subject to} \hspace{0.3cm} &\frac{1}{2}\big[\left\|\mathbf{S}_I\right\|_F^2+\left\|\mathbf{S}_P\right\|_F^2\big]\leq P,\\
&t_0/z_{DC}(\mathbf{S}_P,\mathbf{S}_I,\mathbf{\Phi}_P^{\star},\mathbf{\Phi}_I^{\star},\rho)\leq1,\\
&2^{\bar{R}}/\left[\prod_{n=0}^{N-1}\left(1+\frac{(1-\rho)}{\sigma_n^2} C_n\right)\right]\leq 1.
\end{align}
This is known as a Reverse Geometric Program \cite{Duffin:1973,Chiang:2005}. A similar problem also appeared in the WPT waveform optimization \cite{Clerckx:2015}.
Note that $1/z_{DC}(\mathbf{S}_P,\mathbf{S}_I,\mathbf{\Phi}_P^{\star},\mathbf{\Phi}_I^{\star},\rho)$ and $1/\left[\prod_{n=0}^{N-1}\left(1+\frac{(1-\rho)}{\sigma_n^2} C_n\right)\right]$ are not posynomials, therefore preventing the use of standard GP tools. The idea is to replace the last two inequalities (in a conservative way) by making use of the arithmetic mean-geometric mean inequality.

\par Let $\left\{g_k(\mathbf{S}_P,\mathbf{S}_I,\mathbf{\Phi}_P^{\star},\mathbf{\Phi}_I^{\star},\rho)\right\}$ be the monomial terms in the posynomial $z_{DC}(\mathbf{S}_P,\mathbf{S}_I,\mathbf{\Phi}_P^{\star},\mathbf{\Phi}_I^{\star},\rho)=\sum_{k=1}^K g_k(\mathbf{S}_P,\mathbf{S}_I,\mathbf{\Phi}_P^{\star},\mathbf{\Phi}_I^{\star},\rho)$. Similarly we define $\left\{g_{nk}(\mathbf{S}_I,\rho)\right\}$ as the set of monomials of the posynomial $1+\frac{\bar{\rho}}{\sigma_n^2}C_n=\sum_{k=1}^{K_n}g_{nk}(\mathbf{S}_I,\rho)$ with $\bar{\rho}=1-\rho$. For a given choice of $\left\{\gamma_k\right\}$ and $\left\{\gamma_{nk}\right\}$ with $\gamma_k,\gamma_{nk}\geq 0$ and $\sum_{k=1}^K \gamma_k=\sum_{k=1}^{K_n} \gamma_{nk}=1$, we perform single condensations and write the standard GP as
\begin{align}\label{standard_GP_SWIPT}
\min_{\mathbf{S}_P,\mathbf{S}_I,\rho,\bar{\rho},t_0} \hspace{0.3cm} &1/t_0\\
\textnormal{subject to} \hspace{0.3cm} &\frac{1}{2}\big[\left\|\mathbf{S}_I\right\|_F^2+\left\|\mathbf{S}_P\right\|_F^2\big]\leq P,\\ 
&t_0\prod_{k=1}^K\left(\frac{g_k(\mathbf{S}_P,\mathbf{S}_I,\mathbf{\Phi}_P^{\star},\mathbf{\Phi}_I^{\star},\rho)}{\gamma_k}\right)^{-\gamma_k}\leq1,\\
&2^{\bar{R}}\prod_{n=0}^{N-1}\prod_{k=1}^{K_n}\left(\frac{g_{nk}(\mathbf{S}_I,\rho)}{\gamma_{nk}}\right)^{-\gamma_{nk}}\leq 1,\\
&\rho+\bar{\rho}\leq 1.\label{standard_GP_SWIPT_4}
\end{align}
\par It is important to note that the choice of $\left\{\gamma_k,\gamma_{nk}\right\}$ plays a great role in the tightness of the AM-GM inequality. An iterative procedure can be used where at each iteration the standard GP \eqref{standard_GP_SWIPT}-\eqref{standard_GP_SWIPT_4} is solved for an updated set of $\left\{\gamma_k,\gamma_{nk}\right\}$. Assuming a feasible set of magnitude $\mathbf{S}_P^{(i-1)}$ and $\mathbf{S}_I^{(i-1)}$ and power splitting ratio $\rho^{(i-1)}$ at iteration $i-1$, compute at iteration $i$ $\gamma_k=\frac{g_k(\mathbf{S}_P^{(i-1)},\mathbf{S}_I^{(i-1)},\mathbf{\Phi}_P^{\star},\mathbf{\Phi}_I^{\star},\rho^{(i-1)})}{z_{DC}(\mathbf{S}_P^{(i-1)},\mathbf{S}_I^{(i-1)},\mathbf{\Phi}_P^{\star},\mathbf{\Phi}_I^{\star},\rho^{(i-1)})}$ $k=1,\ldots,K$ and $\gamma_{nk}= g_{nk}(\mathbf{S}_I^{(i-1)},\rho^{(i-1)})/\big(1+\frac{\bar{\rho}^{(i-1)}}{\sigma_n^2}C_n(\mathbf{S}_I^{(i-1)})\big)$, $n=0,\ldots,N-1$, $k=1,\ldots,K_{n}$ and then solve problem \eqref{standard_GP_SWIPT}-\eqref{standard_GP_SWIPT_4} to obtain $\mathbf{S}_P^{(i)}$, $\mathbf{S}_I^{(i)}$ and $\rho^{(i)}$. Repeat the iterations till convergence. The whole optimization procedure is summarized in Algorithm \ref{Algthm_OPT_WIPT}.

\begin{algorithm}
\caption{SWIPT Waveform}
\label{Algthm_OPT_WIPT}
\begin{algorithmic}[1]
\State \textbf{Initialize}: $i\gets 0$, $\bar{R}$, $\mathbf{\Phi}_P^{\star}$ and $\mathbf{\Phi}_I^{\star}$, $\mathbf{S}_P$, $\mathbf{S}_I$, $\rho$, $\bar{\rho}=1-\rho$, $z_{DC}^{(0)}=0$
\label{Algthm_OPT_WIPT_step_initialize}
\Repeat
    \State $i\gets i+1$, $\ddot{\mathbf{S}}_P\gets \mathbf{S}_P$, $\ddot{\mathbf{S}}_I\gets \mathbf{S}_I$, $\ddot{\rho}\gets \rho$, $\ddot{\bar{\rho}}\gets \bar{\rho}$
    \State $\gamma_k\gets g_k(\ddot{\mathbf{S}}_P,\ddot{\mathbf{S}}_I,\mathbf{\Phi}_P^{\star},\mathbf{\Phi}_I^{\star},\ddot{\rho})/z_{DC}(\ddot{\mathbf{S}}_P,\ddot{\mathbf{S}}_I,\mathbf{\Phi}_P^{\star},\mathbf{\Phi}_I^{\star},\ddot{\rho})$, $k=1,\ldots,K$  
    \label{Algthm_OPT_WIPT_step_gamma}
    \State $\gamma_{nk}\gets g_{nk}(\ddot{\mathbf{S}}_I,\ddot{\rho})/\big(1+\frac{\ddot{\bar{\rho}}}{\sigma_n^2}C_n(\ddot{\mathbf{S}}_I)\big)$, $n=0,\ldots,N-1$, $k=1,\ldots,K_{n}$  
    \label{Algthm_OPT_WIPT_step_gamma_2}
    \State  $\mathbf{S}_P,\mathbf{S}_I,\rho,\bar{\rho} \gets \arg \min \eqref{standard_GP_SWIPT}-\eqref{standard_GP_SWIPT_4}$
    \label{Algthm_OPT_WIPT_step_OPT}
    \State $z_{DC}^{(i)} \gets z_{DC}(\mathbf{S}_P,\mathbf{S}_I,\mathbf{\Phi}_P^{\star},\mathbf{\Phi}_I^{\star},\rho)$
\Until{$\left|z_{DC}^{(i)} - z_{DC}^{(i-1)} \right| < \epsilon$ \text{or} $i=i_{\max}$ }
\end{algorithmic}
\end{algorithm}

Similarly to WPT waveform optimization in \cite{Clerckx:2015}, the final solution for the SWIPT waveform optimization problem is not guaranteed to be the global optimum but only a local optimum. 

\section{Simulation Results}\label{simulations}
\par
\begin{figure}
\centerline{\includegraphics[width=0.9\columnwidth]{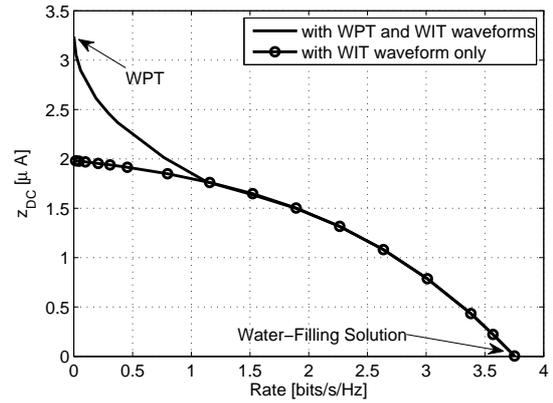}}
  \caption{$C_{R-I_{DC}}$ for $N=16$ and $M=1$.}
  \label{z_DC_results_SWIPT_no_channel}
\end{figure}

 We now illustrate the performance of the optimized SWIPT architecture. $k_2=0.0034$ and $k_4=0.3829$ have been computed for an operating point $a=0$ and used as such to design the optimized waveform. We assume a WiFi-like environment at a center frequency of 5.18GHz with a 36dBm EIRP, 2dBi receive antenna gain and 58dB path loss. This leads to an average received power of about -20dBm. The noise power $\sigma_n^2$ is fixed at -40dBm (i.e.\ 20dB SNR). The frequency gap is fixed as $\Delta_w=2\pi\Delta_f$ with $\Delta_f=B/N$ with $B=1MHz$ and the $N$ sinewaves are centered around 5.18GHz. Fig.\ \ref{z_DC_results_SWIPT_no_channel} illustrates the rate-energy region obtained with Algorithm \ref{Algthm_OPT_WIPT} for $M=1$ and $N=16$ in the particular scenario where the impulse response of the channel is equal to 1. The rate is normalized w.r.t.\ $N$. Extreme points on the x and y-axis refer to the rate and $z_{DC}$ achieved by the water-filling solution and the WPT waveform of \cite{Clerckx:2015}, respectively. The superposition of the WPT and WIT waveforms significantly enlarges the region over the case where WPT waveform is not transmitted (only WIT-OFDM waveform is sent).

\section{Conclusions}\label{conclusions}
The paper derived a methodology to design waveforms for MISO SWIPT. Contrary to the existing SWIPT literature, the non-linearity of the rectifier is modeled and taken into account in the SWIPT waveform and transceiver optimization. The SWIPT waveform is obtained as the superposition of a WPT waveform (multisine) and a WIT waveform (OFDM). The waveforms are adaptive to the CSI (assumed available to the transmitter) and result from a non-convex posynomial maximization problem. The algorithm allows to draw the fundamental limits of SWIPT in terms of rate-energy region. Future interesting works consist in designing SWIPT transceivers for broadcast, multiple access, interference and relay channels accounting for the non-linearity of the rectifier.

\ifCLASSOPTIONcaptionsoff
  \newpage
\fi


\begin{thebibliography}{1}



\bibitem{Zhang:2013} R. Zhang and C. K. Ho, ``MIMO broadcasting for simultaneous wireless information and power transfer,'' IEEE Trans. Wireless Commun., vol. 12, no. 5, pp. 1989-2001, May 2013.

\bibitem{Zhou:2013} X. Zhou, R. Zhang and C. K. Ho, ``Wireless Information and Power Transfer: Architecture Design and Rate-Energy Tradeoff,'' IEEE Trans. on Commun., vol. 61, no. 11, pp. 4754-4767, Nov 2013.

\bibitem{Park:2013} J. Park and B. Clerckx, ``Joint Wireless Information and Energy Transfer in a Two-User MIMO Interference Channel,'' IEEE Trans. Wireless Commun., vol. 12, no. 8, pp. 4210-4221, Aug. 2013.
\bibitem{Park:2014} J. Park and B. Clerckx, ``Joint Wireless Information and Energy Transfer in a K-User MIMO Interference Channel,'' IEEE Trans. Wireless Commun., vol. 13, no. 10, pp. 5781-5796, Oct. 2014.

\bibitem{Huang:2013} K. Huang and E, G. Larsson, ``Simultaneous information and power transfer for broadband wireless systems,'' IEEE Trans. Sig. Process., vol. 61, no. 23, pp. 5972-5986, Dec. 2013.

\bibitem{Nasir:2013} A. A. Nasir, X. Zhou, S. Durrani, and R. A. Kennedy, ``Relaying protocols for wireless energy harvesting and information processing,'' IEEE Trans. Wireless Commun., vol. 12, no. 7, pp. 3622-3636, Jul. 2013.
\bibitem{Huang:2015} Y. Huang and B. Clerckx, ``Joint Wireless Information and Power Transfer for an Autonomous Multiple-Antenna Relay System,'' IEEE Comm. Letters, vol. 19, no. 7, pp. 1113-1116, July 2015.

%
\bibitem{Clerckx:2015} B. Clerckx, E. Bayguzina, D. Yates, and P.D. Mitcheson, ``Waveform Optimization for Wireless Power Transfer with Nonlinear Energy Harvester Modeling,'' IEEE ISWCS 2015, August 2015, Brussels.
\bibitem{Trotter:2009} M.S. Trotter, J.D. Griffin and G.D. Durgin, ``Power-Optimized Waveforms for Improving the Range and Reliability of RFID Systems,'' 2009 IEEE International Conference on RFID.
\bibitem{Boaventura:2011} A. S. Boaventura and N. B. Carvalho, ``Maximizing DC Power in Energy Harvesting Circuits Using Multisine Excitation,'' 2011 IEEE MTT-S International Microwave Symposium Digest (MTT). 
\bibitem{Collado:2014} A. Collado and A. Georgiadis, ``Optimal Waveforms for Efficient Wireless Power Transmission,'' IEEE Microwave and Wireless Components Letters, vol. 24, no.5, May 2014.


\bibitem{Ladan:2015} S. Ladan and K. Wu, ``Nonlinear Modeling and Harmonic Ercycling of Millimeter-Wave Rectifier Circuit,'' IEEE Trans. MTT, vol. 63, no. 3, March 2015.

\bibitem{Clerckx:2013} B. Clerckx and C. Oestges, ``MIMO Wireless Networks: Channels, Techniques and Standards for Multi-Antenna, Multi-User and Multi-Cell Systems,'' Academic Press (Elsevier), Oxford, UK, Jan 2013.


\bibitem{Chiang:2005} M. Chiang, ``Geometric Programming for Communication Systems,'' Foundations and Trends in Communications and Information Theory, 2005.

\bibitem{Duffin:1973} R.J. Duffin and E.L. Peterson, ``Geometric Programming with Signomials,'' Journal of Optimization Theory and Applications, Vol. 11, No. 1, 1973.
\bibitem{CVX} M. Grant, S. Boyd, and Y. Ye, ``CVX: MATLAB software for disciplined convex programming [Online],'' Available: http://cvxr.com/cvx/, 2015.
\end{thebibliography}
\end{document}